\documentclass[usenatbib]{mnras}
\usepackage{amsmath,amssymb}
\usepackage[pdftex]{graphicx}

\def\vb#1{\mbox{\boldmath $#1$}}

\newcommand{\persc}{{\rm cm^{-2}}}
\newcommand{\percc}{{\rm cm^{-3}}}

\newcommand{\E}[1]{\times 10^{#1}}

\newcommand{\nH}{n_{{\rm H}}}
\newcommand{\mH}{m_{{\rm H}}}

\newcommand{\nHmax}{n_{\rm H,max}}
\newcommand{\nHth}{n_{\rm H,th}}

\newcommand{\tSF}{t_{\rm SF}}
\newcommand{\HII}{H~{\sc ii}}

\newcommand{\MJ}{M_{\rm J}}
\newcommand{\lJ}{\lambda_{\rm J}}
\newcommand{\MPopIII}{M_{\rm Pop III}}
\newcommand{\MPopIIIp}{M_{{\rm Pop III},1}}

\newcommand{\tform}{t_{\rm form}}
\newcommand{\tlife}{t_{\rm life}}

\newcommand{\Mmet}{M_{\rm met}}
\newcommand{\Mgas}{M_{\rm gas}}

\newcommand{\QHI}{Q({\rm H})}
\newcommand{\QHeI}{Q({\rm He})}
\newcommand{\QHeII}{Q({\rm He}^+)}
\newcommand{\QLW}{Q({\rm LW})}
\newcommand{\kmpers}{{\rm km \ s^{-1}}}
\newcommand{\cmpers}{{\rm cm \ s^{-1}}}

\newcommand{\cs}{c_{\rm s}}

\newcommand{\aby}[1]{y({\rm {#1}})}
\newcommand{\abyth}[1]{y_{\rm th} ({\rm {#1}})}

\newcommand{\dd}{{\rm d}}
\newcommand{\abn}[1]{n ({\rm {#1}})}
\newcommand{\abN}[1]{N_{\rm {#1}}}
\newcommand{\fsh}{f_{\rm sh}}
\newcommand{\lsh}{l_{\rm sh}}
\newcommand{\lshJ}{l_{\rm sh, J}}
\newcommand{\lshV}{l_{\rm sh, V}}
\newcommand{\lshD}{l_{\rm sh, D}}
\newcommand{\kdiss}{k_{\rm diss}}

\newcommand{\XH}{X_{{\rm H}}}
\newcommand{\Zsun}{{\rm Z_{\bigodot}}}
\newcommand{\Msun}{{\rm M_{\bigodot}}}

\newcommand{\tcomp}{t_{\rm comp}}
\newcommand{\fRT}{f_{\rm RT}}

\newcommand{\svd}{\scalebox{0.5}{$\vdots$}}

\defcitealias{Chiaki15}{C15}
\defcitealias{Chiaki16}{C16}
\defcitealias{Chiaki17}{C17}
\defcitealias{Chiaki18}{C18}
\defcitealias{Chiaki19}{CW19}
\defcitealias{Smith15}{S15}
\defcitealias{Hirano14}{H14}
\defcitealias{Ishigaki14}{I14}
\defcitealias{Tominaga14}{T14}
\defcitealias{Marassi14}{M14}
\defcitealias{Marassi15}{M15}

\usepackage{color}
\definecolor{rev}{rgb}{0.8,0.0,0.0}

\title[Triggered Pop III star formation]
      {Triggered Population III star formation: the effect of H$_2$ self-shielding}

\author[G. Chiaki \& J. H. Wise]
{Gen Chiaki$^{1,2,3}$\thanks{E-mail: gen.chiaki@nao.ac.jp} and
John H. Wise$^{1}$
\\
$^{1}$Center for Relativistic Astrophysics, School of Physics, Georgia Institute of Technology, Atlanta, GA 30332, USA \\
$^{2}$Astronomical Institute, Graduate School of Science, Tohoku University, Aoba, Sendai 980-8578, Japan \\
$^{3}$National Astronomical Observatory of Japan, 2-21-1 Osawa, Mitaka, Tokyo 181-8588, Japan \\
}

\begin{document}

\date{}

\pagerange{\pageref{firstpage}--\pageref{lastpage}} \pubyear{2022}

\maketitle

\label{firstpage}

\begin{abstract}
The multiplicity of metal-free (Population III) stars may influence their feedback efficiency within their host dark matter halos, affecting subsequent metal enrichment and the transition to galaxy formation.
Radiative feedback from massive stars can trigger nearby star formation in dense self-shielded clouds.
In model radiation self-shielding, the H$_2$ column density must be accurately computed.
In this study, we compare two local approximations based on the density gradient and Jeans length with a direct integration of column density along rays.
After the primary massive star forms, we find that no secondary stars form for both the direct integration and density gradient approaches.
The approximate method reduces the computation time by a factor of 2.
The Jeans length approximation overestimates the H$_2$ column density by a factor of 10, leading to five numerically enhanced self-shielded, star-forming clumps.
We conclude that the density gradient approximation is sufficiently accurate for larger volume galaxy simulations, although one must still caution that the approximation cannot fully
reproduce the result of direct integration.
\end{abstract}


%%%%%%%%%%%%%%%%%%%%%%%%%%%%%%%%%%%%%%%%%%%%%
% INTRODUCTION %%%%%%%%%%%%%%%%%%%%%%%%%%%%%%%%%%%
%%%%%%%%%%%%%%%%%%%%%%%%%%%%%%%%%%%%%%%%%%%%%

\section{INTRODUCTION}

The first generation of metal-free (Population III or Pop III) stars are crucial astronomical objects.
Main-sequence Pop III stars emit copious amounts of ultraviolet (UV) photons,
which can have either positive or negative effects on star formation.
Ionizing photons with energies $\geq 13.6$ eV can induce star 
formation in the interstellar medium (ISM) by enhancing the fraction of electrons, 
which catalyze formation reactions of hydrogen molecules \citep{Ricotti01, Johnson06, Yoshida07}.
Dissociating photons in the Lyman-Werner (LW) band ($11.2$--$13.6$ eV)
can suppress star formation by destroying H$_2$ \citep{Stacy12, Hirano15}.
Massive Pop III stars die with supernova (SN) events,
which can also affect star formation \citep{Klein94, Nakamura06, Chiaki13, Magg22}. 
Sufficiently weak explosions can compress the surrounding gas
and trigger star formation in a dense shell.
Strong explosions can completely destroy ambient gas clumps
and suppress star formation.
Additionally, Pop III SNe supply the first 
elements heavier than lithium (metals) and their condensates (dust grains).
Additional cooling from metals and grains can lead the formation of first low-mass stars, 
by inducing the fragmentation of clouds \citep{Omukai00, Schneider03}.
The efficiency of radiation and SN feedback depends not only on the initial
mass function (IMF) of Pop III stars
but also on the number of Pop III stars per host dark matter minihalo (MH).
Even with the state-of-the-art numerical simulations,
it is challenging to predict the IMF and multiplicity within MHs.

Researchers have made valiant efforts to constrain the IMF of Pop III stars
for the past two decades \citep{Bromm99, Abel02, Yoshida03}.
In metal-free collapsing clouds, hydrogen molecular cooling primarily induces fragmentation.
H$_2$ cooling is inefficient
at densities $\gtrsim 10^4~\percc$, where local thermal equilibrium is established.
The mass scale of fragments can be estimated to be the Jeans mass
\begin{equation}
\MJ = 2\E{3} ~\Msun
\left( \frac{\mu}{1.23} \right)^{-3/2}
\left( \frac{\nH}{10^{4}~\percc} \right)^{-1/2}
\left( \frac{T}{200~{\rm K}} \right)^{3/2}
\end{equation}
\citep{Matsuda69}.
Multi-dimensional simulations showed that
the Pop III stellar mass lies in a range of $\sim 10$--$1000~\Msun$ 
\citep{Hirano14, Susa14}. 

The multiplicity of Pop III stars is also crucial for the efficiency of stellar feedback
\citep{Ritter15}.
Recent numerical simulations have shown that, through the fragmentation of turbulent clouds or accretion discs,
binaries and star clusters form at scales of $\sim 10$--$10^4$ AU
\citep{Turk09, Greif12, Susa19, Wollenberg20}.
\citet{Sugimura20} found that hierarchical binary/triplet systems form from an accretion
disc $\sim 10^5$ yr after the formation of the primary protostar.
The primary binary system consists of massive stars ($60$--$70~\Msun$) with a separation
of $\sim 10^4$ AU.
One star hosts a small triplet system with moderately massive companions
($\sim 10~\Msun$) at distances of $10^2$--$10^3$ AU.

In this paper, we focus on another channel of multiple Pop III star formation, so-called ``triggered
star formation'' \citep{Elmegreen77, Whitworth94, Hosokawa05, Hosokawa06}.
After the primary star forms in a MH, it emits ionizing photons
if the star is sufficiently massive ($\gtrsim 10~\Msun$).
%A dense shell (D-type front) is created by large thermal pressure
%in an {\HII} region.
The overpressurized {\HII} region and associated D-type ionization front 
drives a shock wave, creating a dense shell.
%The shell can be a potential star formation site 
It potentially can host star formation if the abundance of
H$_2$ is sufficiently large ($\aby{H_2} \gtrsim 10^{-3}$)
and becomes self-gravitating.
The H$_2$ fraction in the D-type front can be reduced by LW photons emitted by the primary star.
With a sufficiently large column density $\abN{H_2} \gtrsim 10^{14}~\persc$ of hydrogen molecules, 
the self-shielding effect becomes important \citep{Shull78, Federman79},
where the dissociation rate $\kdiss$ is reduced by a shielding factor $\fsh$ 
that is a non-linear function of $\abN{H_2}$ \citep{Draine96}.
Therefore, an accurate estimate of $\abN{H_2}$ is essential to model
star formation, if any, in the D-type front.

It is ideal to calculate the column density by integrating the H$_2$ number density $\abn{H_2}$ 
from the source.
Gas is generally optically thin in the LW band, and photons can propagate farther distance 
than ionizing photons.
Radiation transport is especially difficult and computationally expensive in multi-dimensional simulations.
To save the computational time, approximation methods are often used, where 
the column density is calculated using a typical length scale (shielding length)
defined at each fluid element.
Previous works have often used the length scale associated with the density or velocity gradient.
The former characterizes the length scale of the density structure.
The latter, so-called the Sobolev length, characterizes the length scale where
a molecule cannot absorb redshifted photons \citep{Sobolev60}.
Another common choice is the local Jeans length, which typically characterizes
the size of a collapsing cloud.
Although it should be irrelevant to an expanding {\HII} shell, this approximation is used
in cosmological simulations, where not only radiative feedback but
also star formation take place.

Several groups have studied the validity of the local approximation in
various test problems.
\citet{Greif14} studied the escape probability of H$_2$ line emission
to evaluate the cooling efficiency of Pop III star-forming clouds.
They found that the Sobolev method underestimates the column density because
the scale length of velocity gradient is significantly smaller than the bulk infall velocity
due to turbulent motions.
\citet{WolcottGreen11} and \citet{Hartwig15b} investigated the effect of H$_2$ self-shielding 
of background LW emission in the context of direct-collapse black hole formation.
\citet{WolcottGreen11} found that the Sobolev approach can reproduce the shielding factor
of the direct integration method while
\citet{Hartwig15b} found that the Jeans approach overestimates the LW intensity.
\citet{SafranekShrader17} studied the self-shielding effects of H$_2$ and CO dissociation 
in a galactic disc.
They compared direct ray-tracing and a variety of local approximations: the Jeans, 
Sobolev-length and density-gradient approach.
They showed that the local approximation, especially the Jeans approach,
in contrast with the findings of \citet{Hartwig15b}, can 
reproduce the result of the full ray-tracing calculation well.
In this work, we study the effect of the local approximation
on the efficiency of triggered star formation, by comparing
the full ray-tracing calculation with the density gradient and Jeans length approaches.

Another important numerical parameter is the threshold density $\nHth$ above which star formation
is assumed to occur.
Stars will form if the gas density grows up to
$\gtrsim 10^{19}~\percc$ in a timescale shorter than the dynamical time \citep{Greif12}.
To resolve the gas dynamics in such dense regions, numerical timesteps are
limited by the short Courant timescales ($\lesssim$ yr).
In Mpc-scale cosmological simulations of first-generations of stars and galaxies, 
several authors use $\nHth = 10^5$--$10^7~\percc$
\citep{Smith15, Schauer21}.
Since the D-type front simultaneously contracts and expands with the thermal pressure from the 
inner {\HII} region at comparable timescales,
the density may only tentatively exceed the threshold value if $\nHth$ is too small.
In this work, we will test the convergence for $\nHth = 10^6$ and $10^8~\percc$.

We structure this paper as follows.
In Section \ref{sec:method}, we detail our cosmological hydrodynamics
simulations and the relevant chemical processes.
Then, we describe the results for the different schemes to estimate the
H$_2$ column density and star formation density threshold in Section \ref{sec:results}.
In Section \ref{sec:discussion}, we compare the computational cost
for direct integration and local approximation.
We also discuss the ramifications of the different 
schemes on star formation and feedback in the early Universe.
Finally, we summarize the paper in Section \ref{sec:conclusion}.

Throughout the simulations, we adopt the cosmological parameters $\Omega _{\rm m} = 0.3089$,
$\Omega _{\rm CDM} = 0.2603$, $\Omega _{\Lambda} = 0.6911$, and $H_0 = 67.74 \ {\rm km \ s^{-1} \ Mpc^{-1}}$
\citep{Planck2015}.
We run the simulations in comoving coordinates, but we describe physical quantities in proper coordinates
throughout this paper, unless otherwise specified.
We use the mass fraction of hydrogen nuclei $\XH = 0.76$.
All the figures in this paper are created with 
the {\sc yt} toolkit \citep{yt}.\footnote{\url{https://yt-project.org/}.}

\begin{table*}
\begin{minipage}{\textwidth}
\caption{Initial parameters of each run}
\label{tab:parameters}
(a) Threshold density $\nHth$ for star formation. \\
\begin{tabular}{lccl}
\hline
{\sc enzo} parameter & {\tt n6} & {\tt n8} & Note \\
\hline 
{\tt PopIIIOverDensityThreshold} & {\tt -1e6} & {\tt -1e8} & 
Minimum density for star formation.$^{a}$ \\
\hline
\end{tabular} \\ \\
(b) Approximation methods for LW transfer. \\
\begin{tabular}{lcccl}
\hline
{\sc enzo} parameter & {\tt TestA} & {\tt TestB} & {\tt TestC} & Note \\
\hline 
{\tt RadiativeTransferOpticallyThinH2} & {\tt 0} & {\tt 1} & {\tt 1} & Flag to use local approximation. \\
{\tt RadiativeTransferUseH2Shielding} & {\tt 1} & --- & --- & Flag to calculate the self-shielding function. \\
\hline
{\sc grackle} parameter & {\tt TestA} & {\tt TestB} & {\tt TestC} & Note \\
\hline 
{\tt H2\_self\_shielding} & --- & {\tt 1} & {\tt 3} & Types of local approximation method.$^b$ \\
\hline
\end{tabular}
\medskip \\
Note --- ($a$) If a negative value is assigned, {\sc enzo} uses its absolute value in units of $\percc$. \\
($b$) 
1: Density gradient.
%2: User-supplied length-scale.
3: Jeans length.
\end{minipage}
\end{table*}

%%%%%%%%%%%%%%%%%%%%%%%%%%%%%%%%%%%%%%%%%%%%%
% Method %%%%%%%%%%%%%%%%%%%%%%%%%%%%%%%%%%%%%
%%%%%%%%%%%%%%%%%%%%%%%%%%%%%%%%%%%%%%%%%%%%%
\section{Method}
\label{sec:method}

\subsection{Cosmological simulation}

We run a suite of cosmological simulations with the adaptive mesh refinement
(AMR)/$N$-body simulation code {\sc enzo} \citep{Bryan14, BrummelSmith19}.
We solve the hydrodynamics equations with the piecewise parabolic method (PPM)
in an Eulerian frame \citep{Woodward84, Bryan95}, using
a Harten-Lax-van Leer-Contact (HLLC) Riemann solver to
accurately capture hydrodynamical shocks and compute
advection of chemical species across contact discontinuities.
We follow the DM dynamics with an $N$-body particle-mesh solver
\citep{Efstathiou85, Bryan97}. 

Computational cells are progressively refined by a factor of two in space
when satisfying the following criteria:
\begin{itemize}
\item[(a)] The baryon mass in a cell exceeds $3 m_{\rm b,0} \times 2^{-0.2 L}$ on a refinement level $L$,
where $m_{\rm b,0}$ is the mean baryon mass on the root grid.
\item[(b)] The DM particle mass contained by a cell exceeds $3 m_{\rm dm,0}$,
where $m_{\rm dm,0}$ is the mean DM mass on the root grid.
\item[(c)] The local Jeans length $\lJ$ is resolved less than 64 cells.
\end{itemize}
The negative coefficient $-0.2$ in the exponent of criterion (a) invokes the super-Lagrangian 
refinement for the gas component while criterion (b) 
ensures Lagrangian refinement for the DM.
When the baryon density starts to increase in the run-away collapse phase,
cells are refined mostly on criterion (c).
This criterion warrants that the local Jeans length is resolved sufficiently 
to prevent spurious fragmentation \citep{Truelove97, Turk12}.

We generate the initial conditions in a periodic box with a side length
of $1h^{-1}$ Mpc (comoving) with {\sc music} \citep{Hahn11}.
We initially run a DM-only simulation with a base resolution $512^3$ and identify
the most massive halo with a mass $5.97\E{8} \ \Msun$ at redshift $z = 7$
with a halo-finding code {\sc rockstar} \citep{Behroozi13}.
After initially refining the halo Lagrangian region with two additional AMR levels, i.e., with higher spatial resolution by 
a factor of four, we restart the simulation adding the baryon component.
With this zoom-in strategy, the effective resolution is $2048^3$, and the minimum DM particle
mass is $12.4 \ \Msun$.

\subsection{Pop III star formation}
\label{sec:pop_iii_star_formation}

The main coolant of a primordial cloud is molecular hydrogen. 
To calculate the fraction and cooling rate of H$_2$, we model the non-equilibrium chemistry 
with the chemistry/cooling library {\sc grackle} 
\citep{Smith17, Chiaki19}.\footnote{\url{https://grackle.readthedocs.io/}.}
We solve a chemical network of 15 primordial species, 
e$^-$, H$^+$, H, H$^-$, H$_2^+$, H$_2$,
D$^+$, D, D$^-$, HD$^+$, HD
He, He$^+$, He$^{2+}$ and HeH$^+$.
This chemical network includes 
the collisional ionization/recombination of H/He and
formation/dissociation of H$_2$/HD molecules.
We compute the rates of radiative cooling including
inverse Compton cooling, bremsstrahlung, H/He transition line cooling,
H$_2$ ro-vibrational transition line cooling and HD vibrational transition line cooling.
We also consider chemical heating from
H$_2$ formation, where the binding energy (4.48 eV per molecule) is 
converted to the thermal energy \citep[see][]{Omukai00}.

%In a molecular cloud, 
When certain criteria with a molecular cloud are met, we assume that
a Pop III star forms.
In reality, a star forms after gas is accreted onto a protostellar hydrostatic core 
with a density of $\nH \sim 10^{19} \ \percc$ \citep{Larson69, Greif12}.
In this work, to save the computational cost,
we insert a Pop III star particle in cells that satisfy the following criteria:
\begin{itemize}
\item[(i)] the gas density exceeds a threshold density $\nHth$,
\item[(ii)] the gas flow is convergent, $\nabla \cdot {\vb v} < 0$,
\item[(iii)] the cooling time is less than the dynamical time,
\item[(iv)] the H$_2$ fraction exceeds a threshold value, $\abyth{H_2} = 10^{-3}$.
\end{itemize}
We assign the mass $\MPopIII$ of the star particle, randomly sampling 
from a Larson-type IMF
\begin{equation}
\frac{\dd N}{\dd \log \MPopIII} \propto 
\MPopIII ^{-1.3} \exp \left[ -\left( \frac{M_{\rm char}}{\MPopIII} \right) ^{1.6} \right] ,
\end{equation}
where $M_{\rm char}$ is a characteristic mass of Pop III stars.
We set the minimum, maximum and characteristic mass to $1$, $300$ and
$20~\Msun$, respectively.
%With our random seed, the mass of the primary star is $10.4~\Msun$.
Secondary star formation may be affected by
the structure of {\HII} region created by the primary star, and the
structure will change for different stellar masses.
We test two cases with fixed primary stellar masses of
$\MPopIIIp = 10.4$ and $40.0~\Msun$, called {\tt M10} and {\tt M40}, respectively.

\begin{table*}
%\begin{minipage}{0.84\textwidth}
\begin{minipage}{0.92\textwidth}
\caption{Properties of forming Pop III stars}
\label{tab:stars}
\begin{tabular}{ccccccccccc}
\hline 
$\MPopIIIp$&   $\nHth$ & Test & $\tform$ & $D$      & $\MPopIII$ & $\tlife$ & $\QHI$    & $\QHeI$    & $\QHeII$   & $\QLW$ \\
$[\Msun]$  & $[\percc]$ &      &   [kyr]  & [pc]     & [$\Msun$]  &  [Myr]   & [s$^{-1}$]& [s$^{-1}$] & [s$^{-1}$] & [s$^{-1}$] \\
\hline
  $10.4$   & $10^6$     &  A   & $  0.0$  &    $0.0$ &    $ 10.4$ &  $ 16.9$ & $ 5.42\E{47}$ & $ 1.78\E{47}$ & $ 2.84\E{41}$ & $ 7.96\E{47}$\\
           &            &  B   & $  0.0$  &    $0.0$ &    $ 10.4$ &  $ 16.9$ & $ 5.42\E{47}$ & $ 1.78\E{47}$ & $ 2.84\E{41}$ & $ 7.96\E{47}$\\
           &            &  C   & $  0.0$  &    $0.0$ &    $ 10.4$ &  $ 16.9$ & $ 5.42\E{47}$ & $ 1.78\E{47}$ & $ 2.84\E{41}$ & $ 7.96\E{47}$\\
           &            & \svd & $ 20.6$  &  $0.143$ &    $ 16.0$ &  $  9.6$ & $ 2.04\E{48}$ & $ 7.95\E{47}$ & $ 2.36\E{43}$ & $ 2.78\E{48}$\\
           &            & \svd & $ 30.7$  &  $0.301$ &    $ 28.7$ &  $  5.4$ & $ 9.77\E{48}$ & $ 4.56\E{48}$ & $ 3.54\E{45}$ & $ 1.22\E{49}$\\
           &            & \svd & $ 40.6$  &  $0.734$ &    $ 28.1$ &  $  5.5$ & $ 9.23\E{48}$ & $ 4.29\E{48}$ & $ 2.97\E{45}$ & $ 1.15\E{49}$\\
           &            & \svd & $ 53.9$  &  $0.336$ &    $ 33.3$ &  $  4.8$ & $ 1.40\E{49}$ & $ 6.79\E{48}$ & $ 1.07\E{46}$ & $ 1.71\E{49}$\\
           &            & \svd & $ 63.8$  &  $0.435$ &    $ 18.8$ &  $  8.0$ & $ 3.21\E{48}$ & $ 1.32\E{48}$ & $ 1.04\E{44}$ & $ 4.26\E{48}$\\
\hline                                                 
  $10.4$   & $10^8$     &  A   & $  0.0$  &    $0.0$ &    $ 10.4$ &  $ 16.9$ & $ 5.42\E{47}$ & $ 1.78\E{47}$ & $ 2.84\E{41}$ & $ 7.96\E{47}$\\
           &            &  B   & $  0.0$  &    $0.0$ &    $ 10.4$ &  $ 16.9$ & $ 5.42\E{47}$ & $ 1.78\E{47}$ & $ 2.84\E{41}$ & $ 7.96\E{47}$\\
           &            &  C   & $  0.0$  &    $0.0$ &    $ 10.4$ &  $ 16.9$ & $ 5.42\E{47}$ & $ 1.78\E{47}$ & $ 2.84\E{41}$ & $ 7.96\E{47}$\\
           &            & \svd & $ 37.1$  & $0.0418$ &    $ 16.0$ &  $  9.6$ & $ 2.04\E{48}$ & $ 7.95\E{47}$ & $ 2.36\E{43}$ & $ 2.78\E{48}$\\
           &            & \svd & $ 62.2$  & $0.0390$ &    $ 67.6$ &  $  3.1$ & $ 6.27\E{49}$ & $ 3.47\E{49}$ & $ 8.27\E{47}$ & $ 7.10\E{49}$\\
\hline
  $40.0$   & $10^6$     & A    & $  0.0$  &    $0.0$ &    $ 40.0$ &  $  3.9$ & $ 2.14\E{49}$ & $ 1.08\E{49}$ & $ 3.83\E{46}$ & $ 2.56\E{49}$\\
           &            & B    & $  0.0$  &    $0.0$ &    $ 40.0$ &  $  3.9$ & $ 2.14\E{49}$ & $ 1.08\E{49}$ & $ 3.83\E{46}$ & $ 2.56\E{49}$\\
           &            & C    & $  0.0$  &    $0.0$ &    $ 40.0$ &  $  3.9$ & $ 2.14\E{49}$ & $ 1.08\E{49}$ & $ 3.83\E{46}$ & $ 2.56\E{49}$\\
           &            & \svd & $ 20.3$  &  $0.283$ &    $ 16.0$ &  $  9.6$ & $ 2.04\E{48}$ & $ 7.95\E{47}$ & $ 2.36\E{43}$ & $ 2.78\E{48}$\\
           &            & \svd & $ 30.2$  &  $0.350$ &    $ 28.1$ &  $  5.5$ & $ 9.23\E{48}$ & $ 4.29\E{48}$ & $ 2.97\E{45}$ & $ 1.15\E{49}$\\
           &            & \svd & $ 45.0$  &  $0.680$ &    $ 26.8$ &  $  5.7$ & $ 8.23\E{48}$ & $ 3.77\E{48}$ & $ 2.07\E{45}$ & $ 1.04\E{49}$\\
           &            & \svd & $ 55.0$  &  $0.560$ &    $ 33.3$ &  $  4.8$ & $ 1.40\E{49}$ & $ 6.79\E{48}$ & $ 1.07\E{46}$ & $ 1.71\E{49}$\\
           &            & \svd & $ 66.3$  &   $1.18$ &    $ 18.8$ &  $  8.0$ & $ 3.21\E{48}$ & $ 1.32\E{48}$ & $ 1.04\E{44}$ & $ 4.26\E{48}$\\
           &            & \svd & $ 76.2$  &  $0.704$ &    $ 26.5$ &  $  5.8$ & $ 8.00\E{48}$ & $ 3.65\E{48}$ & $ 1.89\E{45}$ & $ 1.01\E{49}$\\
           &            & \svd & $ 86.2$  &  $0.143$ &    $ 22.1$ &  $  6.8$ & $ 4.97\E{48}$ & $ 2.15\E{48}$ & $ 4.21\E{44}$ & $ 6.43\E{48}$\\
           &            & \svd & $ 96.2$  &  $0.542$ &    $ 14.8$ &  $ 10.5$ & $ 1.62\E{48}$ & $ 6.11\E{47}$ & $ 1.09\E{43}$ & $ 2.23\E{48}$\\
\hline
\end{tabular}
\medskip \\
Note ---
(1) primary Pop III stellar mass.
(2) threshold density for star formation.
(3) ID of tests.
(4) formation time.
(5) distance from the primary star.
(6) mass.
(7) lifetime.
(8--11) emission rates of H, He and He$^+$ ionizing photons and H$_2$ dissociating photons.
\end{minipage}
\end{table*}

\subsection{Radiation feedback from a Pop III star}
\label{sec:radiation_feedback_from_a_pop_iii_star}

During the main sequence of a Pop III star,
we solve the radiative transfer equation with the adaptive ray tracing module {\sc moray} \citep{Wise11}. 
We calculate the number flux $P$ of ionizing/dissociating photons passing
through each computational cell.
From each radiation source, we integrate $P$ along rays in directions
based on {\sc HEALPix} \citep[Hierarchical Equal Area isoLatitude Pixelation;][]{Gorski05}.
The number of rays is $12\times 4^l$ with a level $l$.
The initial level ({\tt RadiativeTransferInitialHEALPixLevel}) 
is set to 1, and rays are adaptively split as they travel away from the source.
We set the minimum number of rays passing through a cell ({\tt RadiativeTransferRaysPerCell})
to 5.1.

We divide the spectral energy distribution of the source into four energy bins of
$(E_{\rm LW},~E_{\rm H},~E_{\rm He},~E_{{\rm He}^+})=(12.8,~28.0,~30.0,~58.0)$ eV, corresponding to the dissociating,
H, He and He$^+$ ionizing photons, respectively.
In each energy bin, we do not consider frequency dependence of the cross-section and
photon flux. 
Instead, we use the averaged values over each energy band to save numerical costs.
The validity of this assumption is discussed in Section \ref{sec:multifrequency}.
Also,
the energies are fixed regardless of the stellar mass for simplicity.
We use the fits from \citet{Schaerer02} to 
calculate the emission rates of dissociating and
H, He, and He$^+$ ionizing photons, $\QLW$, $\QHI$, $\QHeI$ and $\QHeII$,
respectively, as a function of stellar mass.

For ionizing photons, we solve the radiative transfer equation for all runs.
We calculate the optical depth along a ray segment passing through a cell with a size $\dd r$ as
\begin{equation}
\dd \tau _{i} = \sigma _{i} n_i \\d r,
\end{equation}
where $\sigma _{i}$ and $n_i$ is the absorption cross-section
\citep[taken from][]{Verner96} and
number density of a species $i=\{ {\rm H},~{\rm He},~{\rm He}^+ \}$, respectively.
The photon flux is reduced by
\begin{equation}
\dd P_{{\rm ion}, i} = P_{{\rm ion}, i} (1 - e^{-\dd \tau _{i}})
\end{equation}
accross the ray segment.
Then we calculate the photoionization rate as
\begin{equation}
k_{{\rm ion}, i} = \frac{\dd P_{{\rm ion}, i}}{n_i V_{\rm cell} \dd t_{\rm P}}
\end{equation}
during a photon integration timestep $\dd t_{\rm P}$, 
where $V_{\rm cell}$ is a cell volume, from a single ray.
The total photoionization rate is the sum of all the rays passing through the cell. 

To calculate the photodissociation rate, we test the following three methods.

\vspace{0.5cm}
\noindent{{\tt TestA}  {\it Direct integration of H$_2$ number density}}

In this test, we compute
the H$_2$ dissociation rate $\kdiss$ using the number of LW photons
$P_{\rm LW}$ entering a computational cell as
\begin{equation}
\kdiss = \sum _{\rm rays} 
\frac{ P_{\rm LW} \sigma _{\rm H_2} \Omega _{\rm ray} r^2 \dd r }
{ A_{\rm cell} V_{\rm cell} \dd t_{\rm P}},
\end{equation}
where
$A_{\rm cell}$ is the face area of the computational cell,
$r$ is a distance between the source and the cell,
$\Omega _{\rm ray}$ is the solid angle of a {\sc HEALPix} cell.
We use a reaction cross-section $\sigma _{\rm H_2} = 3.71\E{-18}~{\rm cm}^{2}$ of H$_2$
\citep{Abel97}.
In the cell, photons are attenuated as
\begin{equation}
\dd P_{\rm LW} = P_{\rm LW}
\left[
\fsh (\abN{H_2} + \dd \abN{H_2}) - \fsh (\abN{H_2})
\right] ,
\end{equation}
where $\fsh (\abN{H_2})$ is a shielding function.
We use the fitting function
\begin{eqnarray}
\fsh (\abN{H_2}) &=& \frac{0.965}{(1 + x/v_{\rm th,5})^2} + \frac{0.035}{(1+x)^{0.5}} \nonumber \\
&& \times \exp \left[ -8.5\E{4} (1+x)^{0.5} \right] 
\label{eq:fsh}
\end{eqnarray}
\citep{WolcottGreen11},
by setting the parameter {\tt RadiativeTransferH2ShieldType = 1},
where $x = \abN{H_2} / 5\E{14}~\persc$ and
$v_{\rm th,5} = v_{\rm th} / 10^5~\cmpers $.
The column density is directly integrated as
\begin{equation}
\abN{H_2} = \int \abn{H_2} \dd s
\label{eq:direct}
\end{equation}
along a {\sc HEALPix} ray $s$.

\vspace{0.5cm}
\noindent{{\tt TestB}  {\it Local approximation with the density gradient }}

In local approximations, the dissociation rate is estimated as
\begin{equation}
\kdiss = \fsh (\abN{H_2})
\frac{\QLW \sigma _{\rm H_2}}{4\pi r^2}
\end{equation}
using the functional form $\fsh (\abN{H_2})$ given in Eq. (\ref{eq:fsh}).
We compute the column density as 
\begin{equation}
\abN{H_2} = \abn{H_2} \lsh
\label{eq:local}
\end{equation}
with a length scale $\lsh$ (shielding length) defined with physical quantities
of each fluid element.
In {\tt TestB}, we estimate $\lsh$ as the density gradient
\begin{equation}
\lshD = \frac{\rho}{\left|  \nabla \rho \right|},
\end{equation}
where $\rho$ is the density of a cell.

\vspace{0.5cm}
\noindent{{\tt TestC} {\it Local approximation with the Jeans length}}

In this test, we use the same local approximation as {\tt TestB} (Eq. \ref{eq:local}), but
the shielding length is calculated from the local Jeans length

\begin{equation}
\lshJ \equiv \lJ = \left( \frac{\pi \cs^2}{G \rho } \right)^{1/2},
\label{eq:jeans}
\end{equation}
where $\cs$ is the sound speed of a cell.

\begin{figure*}
\includegraphics[width=\textwidth]{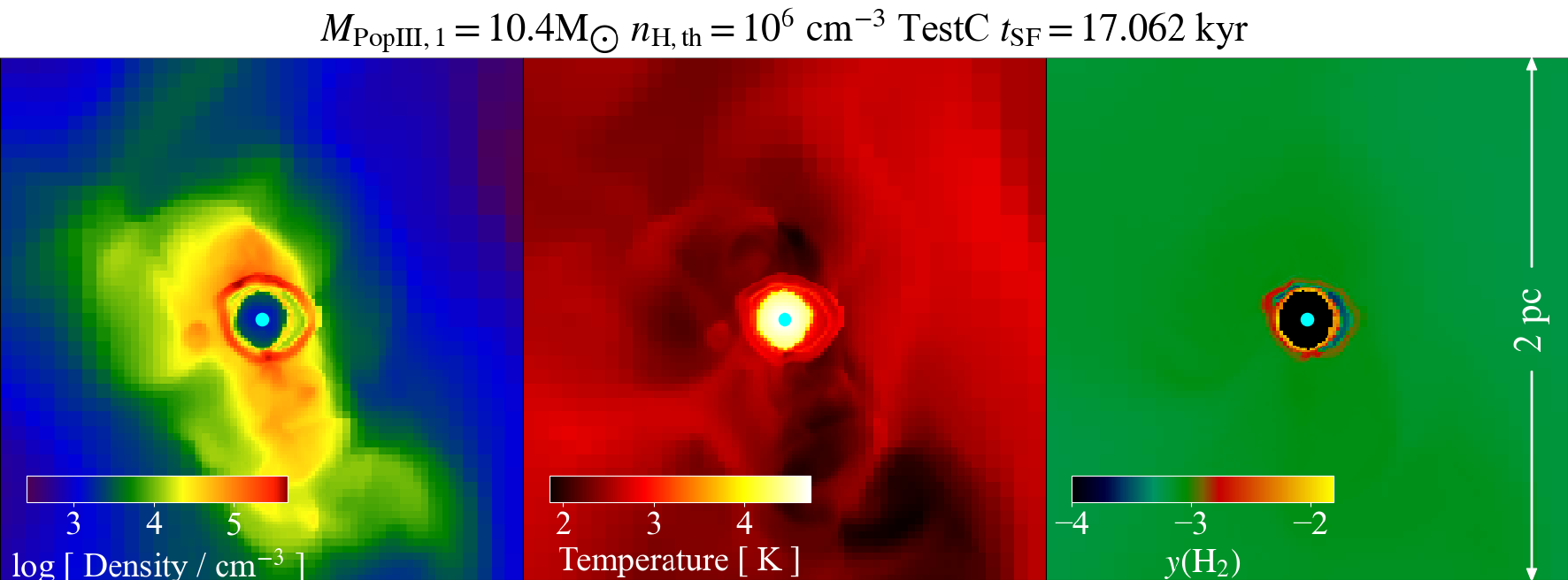}
\caption{
Slices of (a) density, (b) temperature and (c) number fraction $\aby{H_2}$ of
hydrogen molecules relative to hydrogen nuclei for {\tt TestC}
at the time 17.1 kyr after the primary star formation and just before the secondary
star formation.
The plotted window is centered at the position
of the primary Pop III star (cyan circle) with a side length of 2 pc
on the computational $x$-$y$ plane.}
\label{fig:snapshots_1stP3}
\end{figure*}

\begin{figure*}
\includegraphics[width=0.8\textwidth]{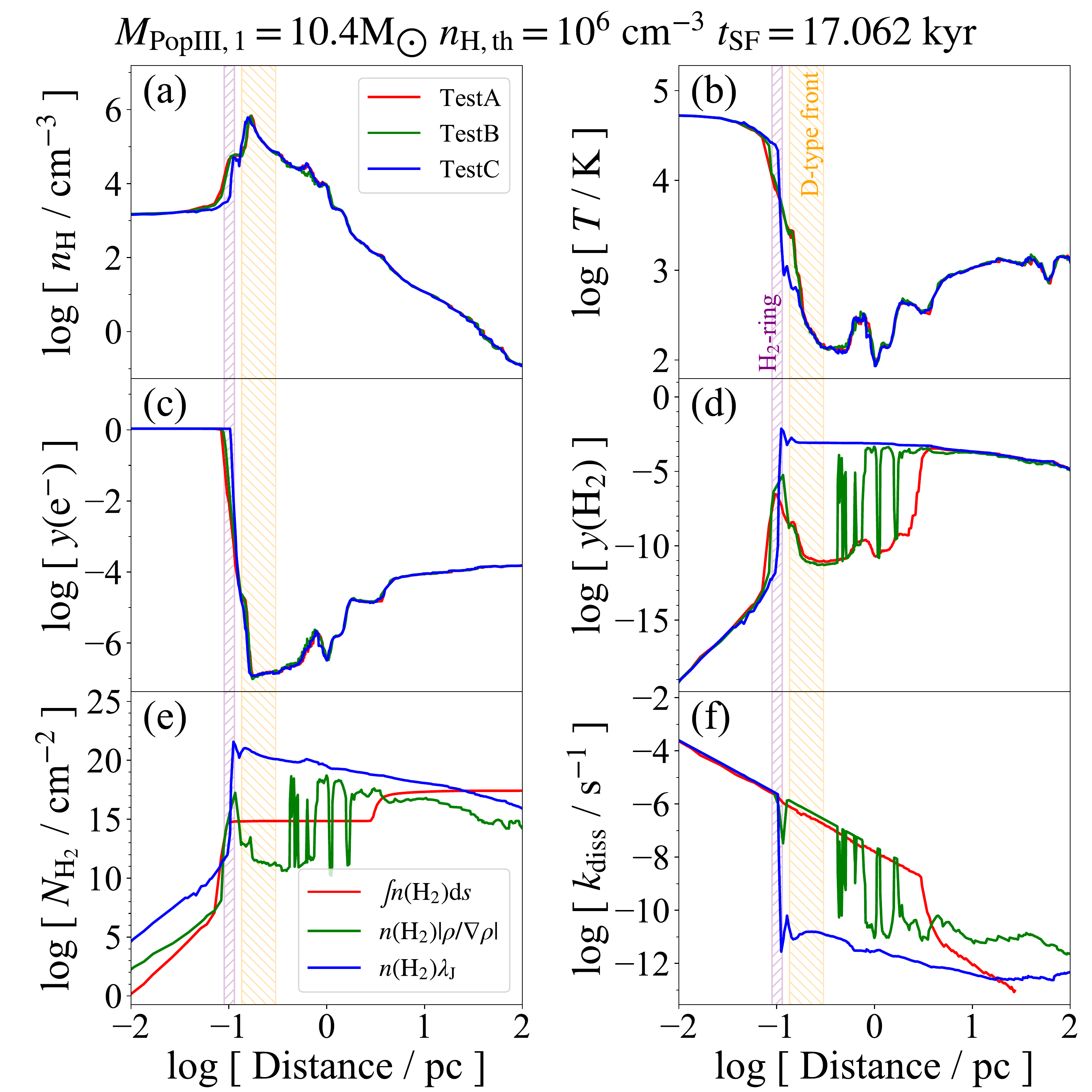}
\caption{
(a) Density $\nH$, (b) temperature $T$, (c) e$^-$ fraction $\aby{e^-}$ (solid curves), 
H$^-$ fraction $\aby{H^-}$ (dashed curves), (d) H$_2$ fraction $\aby{H_2}$, 
(e) H$_2$ column density $\abN{H_2}$ and (f) dissociation rate $\kdiss$ 
as a function of distance from the primary Pop III star 
on a ray from the source to the density maximum
at the time 17.1 kyr after the primary star formation.
The red, green and blue curves denote the results for {\tt TestA}, {\tt TestB} and {\tt TestC},
respectively.
The purple and orange hatched regions represent the H$_2$-ring and D-type front for {\tt TestA}, respectively.
}
\label{fig:rad_1stP3_0031}
\end{figure*}

\begin{figure}
\includegraphics[width=\columnwidth]{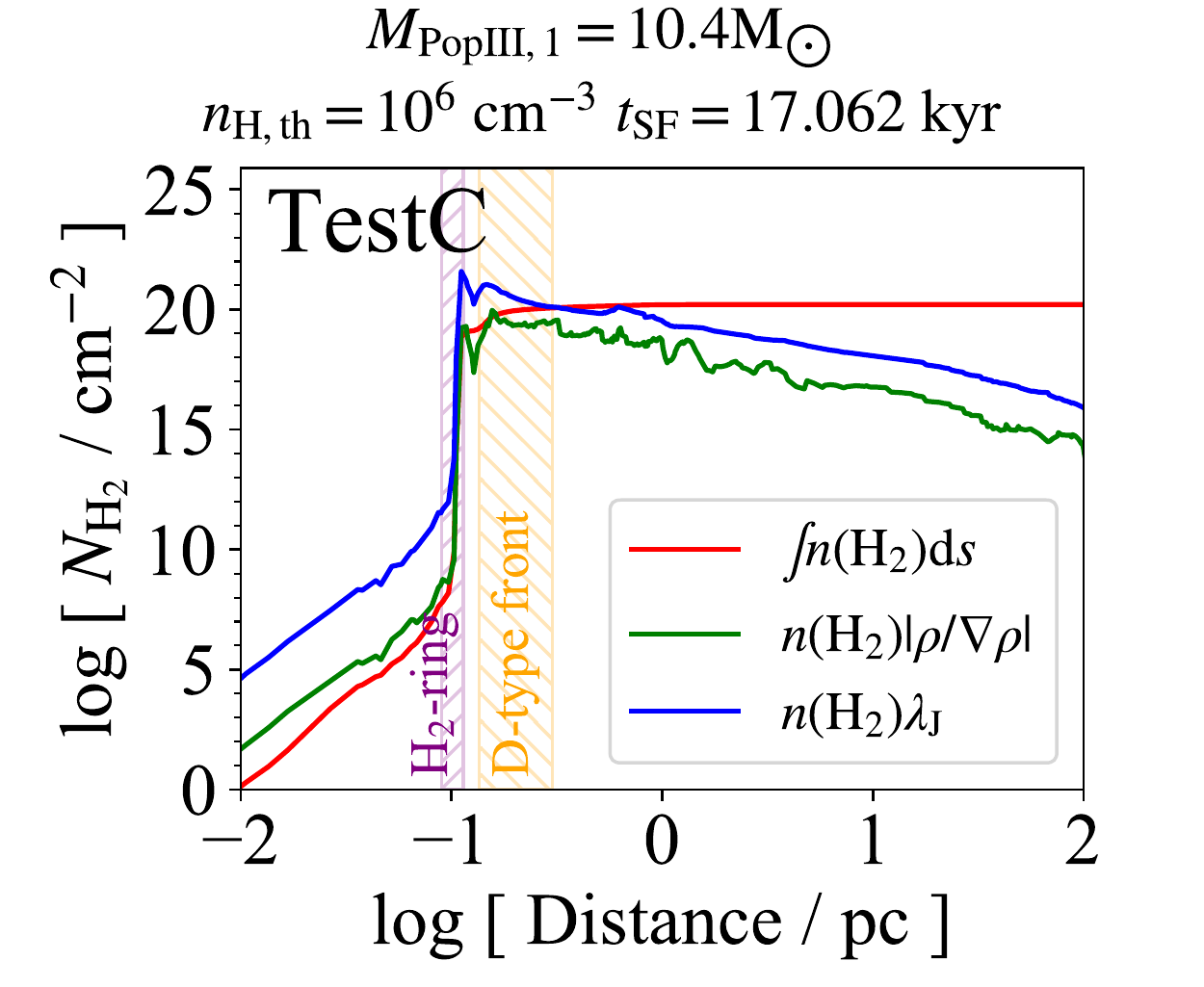}
\caption{
H$_2$ column density calculated with different schemes
as a function of distance from the primary Pop III star 
on a ray from the source to the density maximum
at the time 17.1 kyr after the primary star formation.
From the snapshot of {\tt TestC}, we calculate the column density
by integrating H$_2$ density along the ray (red curve) and
using density gradient $\left| \rho / \nabla \rho \right|$ (green curve).
The blue curve is the same as the one in Fig. \ref{fig:rad_1stP3_0031}e.
The purple and orange hatched regions represent the H$_2$-ring and D-type front, respectively,
the same as Fig. \ref{fig:rad_1stP3_0031}.
}
\label{fig:rad_1stP3_0031_TestC}
\end{figure}

\begin{figure*}
\includegraphics[width=0.8\textwidth]{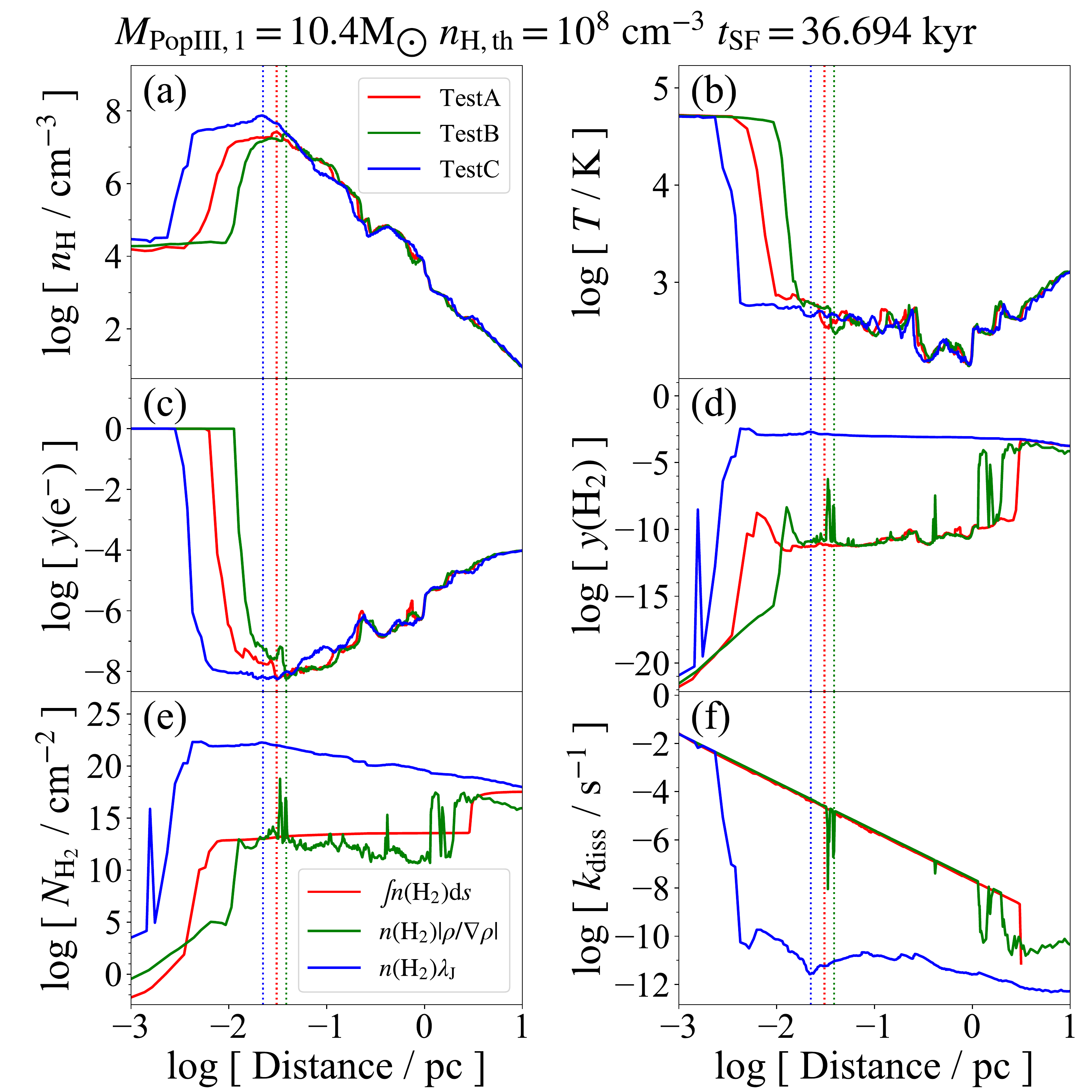}
\caption{
Same as Fig. \ref{fig:rad_1stP3_0031} but at the time 36.7 kyr after the primary
Pop III star formation just before the secondary star formation for {\tt TestC}
for the threshold density $\nHth = 10^8~\percc$ for star formation.
The vertical dotted lines show the distance of the density maxima.
The red, green and blue curves indicate the results for {\tt TestA}, {\tt TestB}
and {\tt TestC}, respectively.
}
\label{fig:rad_1stP3h_0035}
\end{figure*}

\begin{figure}
\includegraphics[width=\columnwidth]{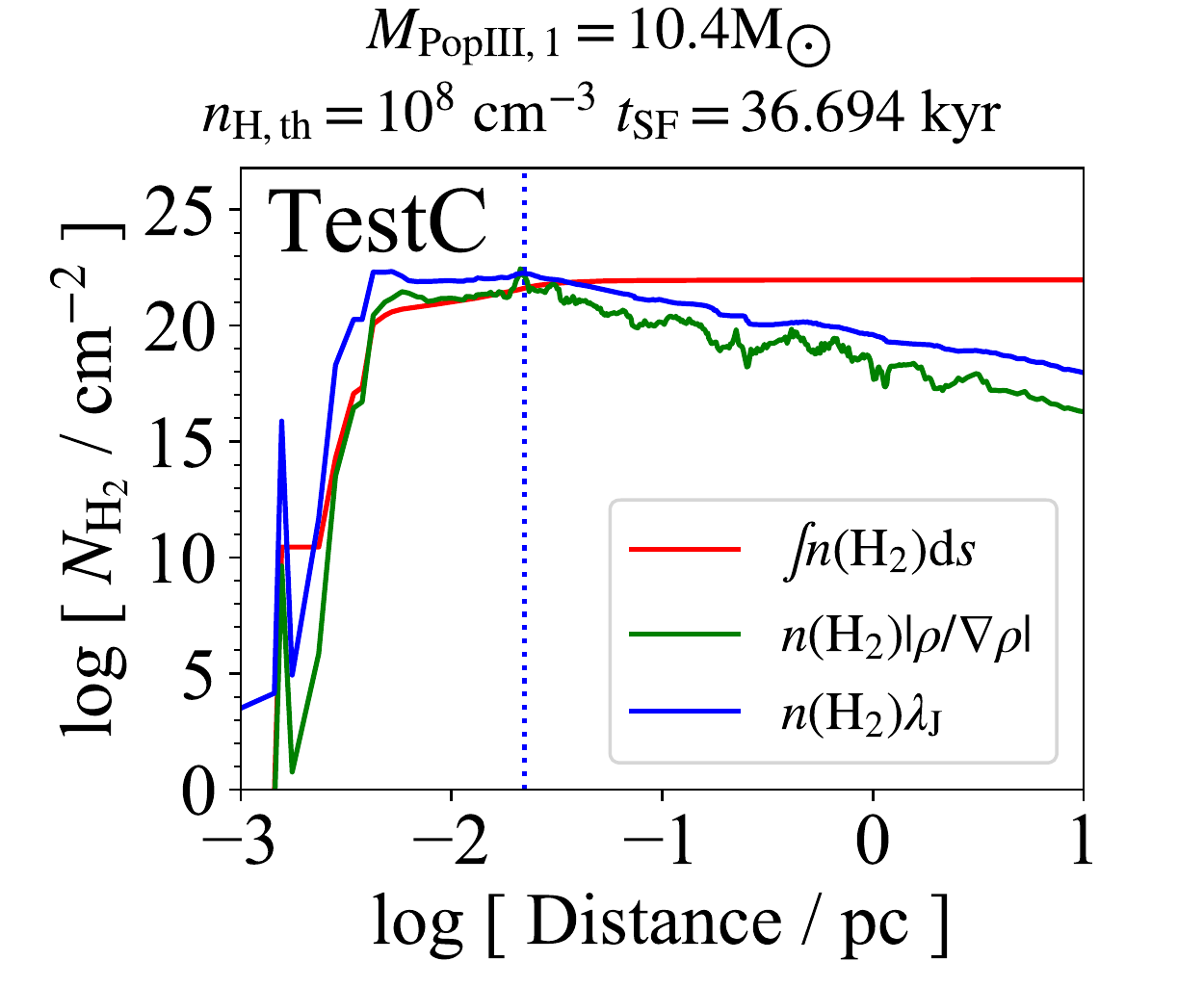}
\caption{
The same as Fig. \ref{fig:rad_1stP3_0031_TestC} but at the time 36.7 kyr after the primary
Pop III star formation just before the secondary star formation for {\tt TestC}
for the threshold density $\nHth = 10^8~\percc$ for star formation.
The vertical dotted line shows the distance of the density maximum.
}
\label{fig:rad_1stP3h_0035_TestC}
\end{figure}

\begin{figure*}
\includegraphics[width=0.8\textwidth]{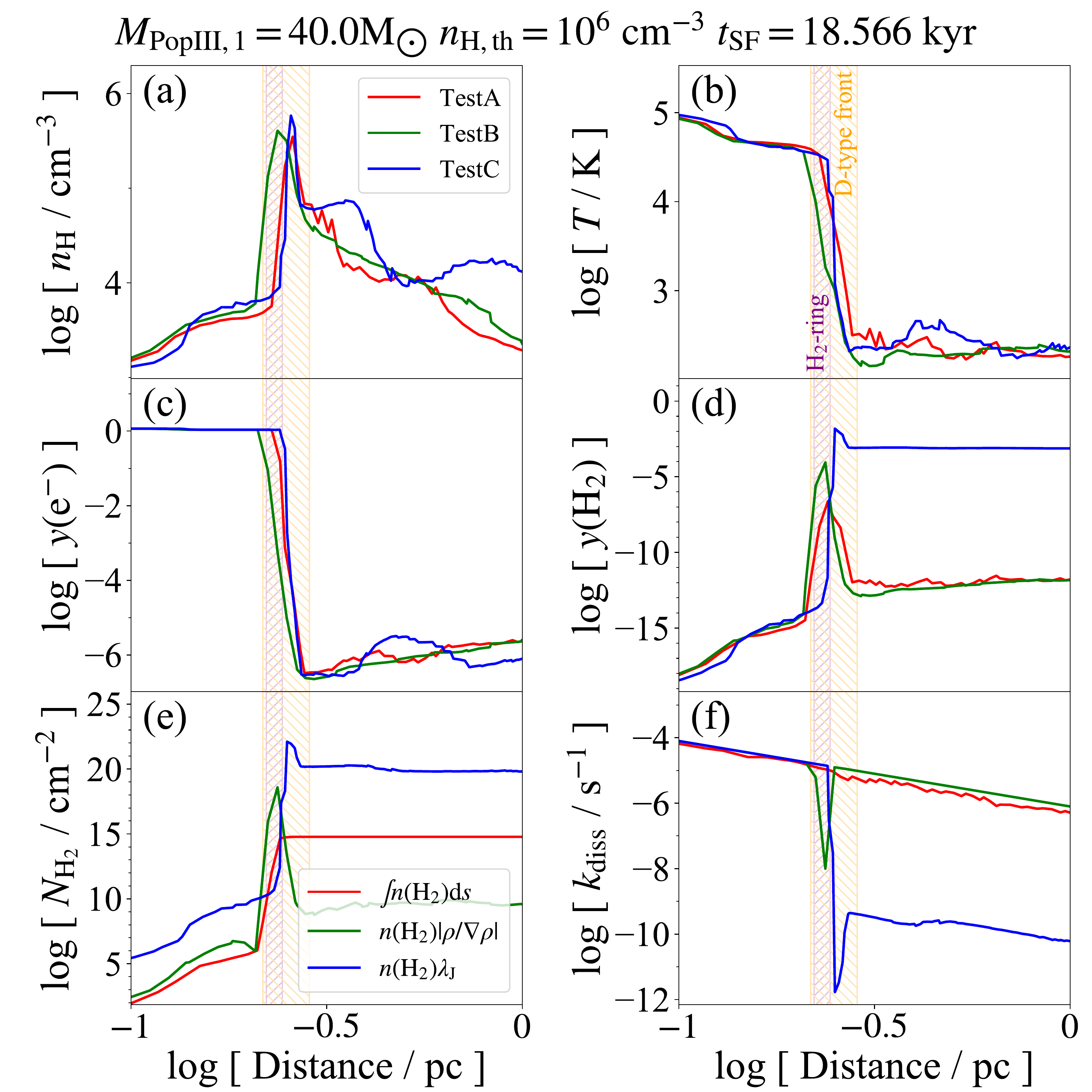}
\caption{
Same as Fig. \ref{fig:rad_1stP3_0031} but at the time 18.6 kyr after the primary
Pop III star formation just before the secondary star formation for {\tt TestC}
for the primary Pop III stelalr mass of $40~\Msun$.
The purple and orange hatched regions represent the H$_2$-ring and D-type front for {\tt TestB}, respectively.
}
\label{fig:rad_1stP3m_0031}
\end{figure*}

\subsection{Star formation density threshold}

The threshold density $\nHth$ can affect the efficiency of Pop III star formation.
For {\tt M10},
we compare two cases with $\nHth = 10^6$ and $10^8~\percc$, 
hereafter called {\tt n6} and {\tt n8}, respectively.
The former value is often used in small-volume cosmological simulations
of the first galaxies \citep{Skinner20, Schauer21} and lower values in larger-volume
simulations \citep{Wise12, Xu16, Jeon21}.
For {\tt M40}, we test only
the case with $\nHth = 10^6~\percc$.
In this paper, we mainly show the result for {\tt M10n6} as a fiducial case.
We run simulations of {\tt TestA}, {\tt B} and {\tt C} for each $\nHth$, and
Table \ref{tab:parameters} summarizes the initial parameters in the runs.

We terminate our simulations 0.1 Myr after the formation of
the primary star,
that is shorter than the lifetime ($\sim 10$ Myr) of a star with 
a mass $\sim 10~\Msun$.
We confirm that the number of Pop III stars is unchanged
by running the simulation for {\tt TestB} until the lifetime of the primary star.
We output snapshots at every 5000 yr to analyze the star formation
history.
Hereafter, we measure the time $\tSF$ from the primary star formation.

%%%%%%%%%%%%%%%%%%%%%%%%%%%%%%%%%%%%%%%%%%%%%
% Results %%%%%%%%%%%%%%%%%%%%%%%%%%%%%%%%%%%%%
%%%%%%%%%%%%%%%%%%%%%%%%%%%%%%%%%%%%%%%%%%%%%
\section{Results}
\label{sec:results}

\subsection{Number of forming Pop III stars}

In this section, we present the results for {\tt M10n6}.
%, where the threshold density $\nHth$ for star formation is $10^6~\percc$. 
At a redshift of $z=25.1$
in a MH with a virial mass of $3.39\E{5}~\Msun$ and virial radius of $86.8$ pc,
H$_2$ molecules form through the reactions
\begin{eqnarray}
{\rm H} + {\rm e}^-  &\to & {\rm H}^- + \gamma ~~~({\rm Reaction~7}), \nonumber \\
{\rm H}^- + {\rm H} & \to & {\rm H}_2 + {\rm e}^- ~~({\rm Reaction~8}), \nonumber
\end{eqnarray}
catalyzed by free electrons \citep[H$^-$-process;][]{Peebles80}.
The gas temperature decreases below $\sim 1000$ K through ro-vibrational
cooling of H$_2$. 
A cloud collapses in a run-away manner.
When the density reaches $10^6~\percc$, we insert 
the primary Pop III star.
With our random seed, the mass of the primary star is assigned to $10.4~\Msun$.
Its lifetime is $\tlife = 16.9$ Myr, and the photon emission
rates are $(\QLW, \QHI, \QHeI, \QHeII) = 
(5.42\E{47}, 1.78\E{47}, 2.84\E{41}, 7.96\E{47}) ~{\rm s}^{-1}$.
An {\HII} region forms around the star through the absorption of ionizing
photons, and modifies the density structure in the ISM.
Photons in the LW band dissociate H$_2$ molecules and affect
the formation of any secondary star.

The subsequent star formation history varies for different 
local approximation methods.
Table \ref{tab:stars} summarizes the properties of Pop III stars forming
during the first 0.1 Myr after the primary star forms.
Only one Pop III star forms for {\tt TestA} and {\tt B},
while six stars form for {\tt TestC}.
In {\tt TestC}, the secondary stars with (random) masses $20$--$30~\Msun$ form
at distances $D= 0.1$--$0.7$ pc from the primary star at the time $\tSF = 20$--$60$ kyr.
Compared to {\tt TestA}, the number of Pop III stars is consistent for {\tt TestB} and 
overestimated for {\tt TestC}.

In the following subsections, we 
describe the evolution of the {\HII} region for {\tt TestA} and {\tt B} (Section \ref{sec:ionizing_region})
and interpret the secondary star formation in case {\tt TestC} (Section \ref{sec:TestC}).
Then, we describe the result for {\tt n8} in Section \ref{sec:n8}.

\subsection{Evolution of {\HII} regions}
\label{sec:ionizing_region}

Fig. \ref{fig:snapshots_1stP3} shows the slices of density,
temperature and the number fraction $\aby{H_2}$ of H$_2$ to hydrogen
nuclei at $\tSF = 17.1$ kyr, just before the secondary star formation for {\tt TestC}.
In this figure, we plot the results only for {\tt TestC}.
The distribution of density and temperature for {\tt TestA} and {\tt B}
is almost the same as {\tt TestC}, but $\aby{H_2}$ is smaller than
for the two other tests by eight orders of magnitude in the dense shell outside the {\HII} region.

Fig. \ref{fig:rad_1stP3_0031} shows the density,
temperature, e$^-$ and H$_2$ fraction, column density and
photodissociation rate along a ray from the source to the density maximum.
The UV photons with energies $E \geq 13.6$ eV emitted by the primary star
ionizes the adjacent gas.
The temperature increases to $\sim 5\E{4}$ K, comparable to 
the surface temperature of the star.
Due to the strong thermal pressure ($\sim 10^{-9}~{\rm dyn/cm^{-2}}$),
the density declines to $\sim 400~\percc$ in the ionized region.
Just outside the ionizing front (I-front), a dense shell (D-type front)
forms.
We define the D-type front as the region with densities above 0.1 times
the maximum density (orange hatched region in Fig. \ref{fig:rad_1stP3_0031}).
In the region between the I-front and D-type front, 
the gas is partly ionized with an electron fraction of $\aby{e} \sim 10^{-3}$.
The H$_2$ fraction increases through the H$^-$-process (Reactions 7 and 8), and
a so-called ``H$_2$-ring'' appears (Fig. \ref{fig:snapshots_1stP3}c).
We define the H$_2$-ring as the region with H$_2$ fractions above 0.1 times
the maximum, $y_{\rm max} ({\rm H}_2) \sim 10^{-7}$ (purple hatched region in Fig. \ref{fig:rad_1stP3_0031}).
In Fig. \ref{fig:rad_1stP3_0031}, we plot the D-type front and H$_2$-ring for {\tt TestB},
but their positions are almost the same for the other tests.

At the time $\tSF = 17.1$ kyr, the radius of the D-type front reaches $\sim 0.1$ pc
(Fig. \ref{fig:rad_1stP3_0031}).
The gas flow is convergent, and the density
increases up to $\sim 10^6~\percc$, comparable to $\nHth$.
Therefore, the D-type front naturally satisfies the criteria (i) and (ii) for
star formation.
If H$_2$ fraction is larger than $\abyth{H_2}$,
the criteria (iii) and (iv) will be also satisfied.
For {\tt TestA} and {\tt B}, due to ineffective shielding,
a sufficient fraction of dissociating photons can 
penetrate into the D-type front, and the formation of
secondary stars is prevented.
We discuss the result in a quantitative manner in subsequent sections.

\vspace{0.3cm}
\noindent{(a) {\tt TestA}}

For {\tt TestA}, the density reaches the maximum value of 
$\nHmax = 6.88\E{5}~\percc$ in the D-type front at a distance $0.170$ pc
at the time $\tSF = 17.1$ kyr.
Since $\nHmax$ is comparable to $\nHth$, the criteria (i) and (ii) will be satisfied
if the convergence continues.
Just inside the D-type front, the H$_2$-ring forms at a distance $0.0960$ pc,
where the fraction of H$_2$ reaches only up to $\aby{H_2} = 2.95\E{-7}$.
The column density increases rapidly in the H$_2$ ring, and it reaches a plateau of
$\abN{H_2} \sim 7\E{14}~\percc$ (red curves in Fig. \ref{fig:rad_1stP3_0031}).
At the density maximum, the column density is $\abN{H_2} = 6.98\E{14}~\persc$.
With the temperature $856$ K, the shielding fraction is $\fsh = 0.704$ (Eq. \ref{eq:fsh}), that is,
dissociation photons are only marginally shielded in the H$_2$-ring.
Since H$_2$ molecules cannot avoid dissociation,
the H$_2$ fraction declines down to $\aby{H_2} \sim 10^{-10}$.
The star-formation criteria (iii) and (iv) are not satisfied, and thus secondary stars
do not form.

\vspace{0.3cm}
\noindent{(b) {\tt TestB}}

For {\tt TestB}, only one Pop III star forms during the simulation,
which is the same result as {\tt TestA}.
However, this does not necessarily means that the Sobolev-like approximation can
reproduce the result of direct integration.
The D-type front is a potential star forming site, because its density ($6.74\E{5}~\percc $) is 
comparable to $\nHth$ at $\tSF = 17.1$ kyr.
The shielding length $\lshD = 0.111$ pc at the density maximum
characterizes the length scale of the D-type front ($0.139$ pc).
The column density is estimated as the product of H$_2$ fraction at the density maximum and
the thickness of the D-type front.
As we have discussed in the previous section, 
the H$_2$-ring mostly contributes to
the column density at the density maximum for {\tt TestA}.
This indicates that the local approximation fails to capture the contribution of
the spatially separated region.
The column density is indeed underestimated to be $\abN{H_2} = 2.95\E{13}~\persc$,
compared to the value $6.98\E{14}~\percc$ for {\tt TestA} by a factor of 20,
because the H$_2$ fraction is smaller in the D-type front than in the H$_2$-ring.
Nevertheless, since the shielding factor $\fsh (\abN{H_2})$ is insensitive to $\abN{H_2}$
at $\abN{H_2} \lesssim 5\E{14}~\percc$, $\fsh = 0.982$ is comparable to the value for
{\tt TestA}.
%The H$_2$ fraction $1.27\E{-10}$ is smaller than the threshold value, and
Consequently, secondary stars do not form as for {\tt TestA}.

\subsection{Secondary star formation for {\tt TestC}}
\label{sec:TestC}

For {\tt TestC}, the shielding factor is overestimated, compared to 
{\tt TestA}, that consequently overproduces Pop III stars.
At $\tSF = 17.1$ kyr, just before the formation of the first secondary star,
the dissociation rate $\kdiss = 8.48\E{-12}~{\rm s}^{-1}$
is much smaller than for {\tt TestA} and {\tt B} at the density maximum.
A large fraction ($9.67\E{-4}$) of H$_2$ survives, 
and $\aby{H_2}$ exceeds the threshold value $10^{-3}$ at $\tSF = 20.6$ kyr.
Feedback from the secondary stars further induces repetitive
star formation.
By $\tSF = 0.1$ Myr, six Pop III stars form.

This behavior occurs because the Jeans length approach
overestimates the H$_2$ column density.
For a fair comparison, we calculate $\abN{H_2}$ with the three different methods
from a snapshot for {\tt TestC} at the time $\tSF = 17.1$ kyr (Fig. \ref{fig:rad_1stP3_0031_TestC}).
The blue and green curves show the result for the direct integration and
the density gradient approach, respectively.
The Jeans length approach overestimates $\abN{H_2} = 9.45\E{20}~\persc$ at the density maximum, 
compared to $4.87\E{19}$ and $9.19\E{19}~\persc$ for direct integration and
density gradient approach by a factor of two and ten, respectively.
Since the Jeans length originally characterizes the length scale of a quasi-static collapsing cloud,
it is larger than the length scale of the D-type front contracting with the thermal pressure
of the {\HII} region.
The shielding factor $\fsh = 8.36\E{-6}$ is smaller than the direct integration method,
because
$\fsh (\abN{H_2})$ is a decreasing function of $\abN{H_2}$ for $\abN{H_2} \gtrsim 5\E{14}~\percc$ (Eq. \ref{eq:fsh}).
Dissociating photons cannot penetrate into the D-type front due to the high efficiency
of self-shielding, and thus the H$_2$ fraction exceeds the critical value.
After this point, secondary stars form in the D-type front.

\subsection{Effect of the threshold density for star formation}
\label{sec:n8}

%In the previous sections, we have described the result for {\tt n6}.
%Star particles spawn in the D-type front when the density exceeds $\nHth$ for {\tt TestC}.
%This is an artificial setup often used in large-volume cosmological simulations
%to cap dense regions that limit the computational timestep with a short Courant time.
%Without the procedure, the density might decrease by the strong thermal pressure 
%from the radiation source before stars indeed form.
In this section, we describe the results for a higher threshold density 
$\nHth = 10^8~\percc$ ({\tt n8}).
The primary star forms at a redshift $25.0$, 0.3 Myr later than {\tt n6}.
The stellar UV photons creates an {\HII} region, and a D-type front forms just outside
a H$_2$-ring like the {\tt n6} case.
As Table \ref{tab:stars} summarizes,
only one Pop III star forms in {\tt TestA} and {\tt B}, while
three stars form in {\tt TestC} when we terminate the simulations at the time $\tSF = 0.1$ Myr.
For {\tt TestC}, the number of Pop III stars becomes smaller than case {\tt n6}
and approaches the value for {\tt TestA}.
However, the result still does not converge
even for the high $\nHth = 10^8~\percc$.

Fig. \ref{fig:rad_1stP3h_0035} shows density, temperature, e$^-$ and H$_2$ fraction, 
H$_2$ column density and dissociation rate along
a ray from the primary star to the density maximum at $\tSF = 36.7$ kyr,
just before the secondary star formation for {\tt TestC}.
For {\tt TestA}, the column density increases only up to $1.43\E{13}~\percc$
in the H$_2$-ring at a distance $6.29\E{-3}$ pc.
The H$_2$-ring is optically thin ($\fsh = 0.987$), and almost all LW photons
enter the D-type front at a distance $0.0308$ pc (red vertical line in Fig. \ref{fig:rad_1stP3h_0035}).
The H$_2$ fraction is $7.22\E{-12}$, well below the threshold for star formation.

For {\tt TestB}, the shielding length is $\lshD = 0.151$ pc, and
$\abN{H_2} = 4.68\E{16}~\percc$ at the density maximum 
(green dotted line in Fig. \ref{fig:rad_1stP3h_0035}).
The corresponding shielding factor is small ($0.0192$), but 
$\abN{H_2}$ just increases temporarily.
Around the density maximum, we can see spikes in $\abN{H_2}$
with a height of six orders of magnitude 
and a width of $\sim 3\E{-3}$ pc 
(green curve in Fig. \ref{fig:rad_1stP3h_0035}e).
The sound crossing time of the spikes is $t_{\rm sc} \sim 1$ kyr
for the temperature $300$ K,  shorter than the dynamical time $\sim 10$ kyr.
This indicates that the spikes are dumped very quickly.
The small-scale noise is generated
by the sensitivity of the length $\lshD$ to the turbulent motion (convergent flow)
of the gas.
In the D-type front, the column density is $\sim 10^{13}~\persc$
on average, which corresponds to $\fsh \sim 1$.
Therefore, nearly all of the H$_2$ molecules are destroyed by LW photons, and further
star formation does not occur in the D-type front.

For {\tt TestC}, the shielding length is $\lshJ = 0.0392$ pc, and
the column density is $\abN{H_2} = 1.74\E{22}~\persc$.
As in Section \ref{sec:TestC},
we compare column densities calculated with the three different methods, 
using a snapshot for {\tt TestC} at the time $\tSF = 36.7$ kyr 
(Fig. \ref{fig:rad_1stP3h_0035_TestC}).
The column density is overestimated with respect to $4.12\E{21}~\percc$ 
for the direct integration by a factor of four.
The shieling factor is $\fsh = 5.44\E{-8}$, and $\kdiss$ in the D-type
front is small ($2.70\E{-12}~{\rm s}^{-1}$).
As a result, the criterion for star formation is satisfied, and
the secondary stars form in this test.

\subsection{Effect of the primary stellar mass}

For {\tt M40},
the result is the same as {\tt M10}:
no additional stars form for {\tt TestA} and {\tt TestB} while 
8 additional stars form in a D-type front for {\tt TestC} (Table \ref{tab:stars}).
Fig. \ref{fig:rad_1stP3m_0031} shows the radial profiles of the physical values 
just before the secondary star formation for {\tt TestC}.
For {\tt TestC}, the H$_2$ fraction is $9.25\E{-3}$ at the density maximum,
which is by six orders of magnitude larger than $4.29\E{-9}$ for {\tt TestA} (Fig. \ref{fig:rad_1stP3m_0031}d).
This is because the column density is significantly overestimated to be $8.83\E{21}~\persc$,
compared to the value $5.92\E{14}~\persc$ for {\tt TestA} (Fig. \ref{fig:rad_1stP3m_0031}e).
Since the shielding factor is underestimated, the flux of dissociation photons is smaller for {\tt TestC}
(Fig. \ref{fig:rad_1stP3m_0031}f).

For {\tt TestB}, the result is the same as {\tt M10}, 
but the shielding factor is moderately underestimated
at the density maximum, opposite to {\tt M10}.
For {\tt M40}, the H$_2$-ring (purple hatched regions in Fig. \ref{fig:rad_1stP3m_0031}) 
overlaps with the D-type front (orange hatched regions), 
because the structures are radially contracted more due to
the stronger thermal pressure in the {\HII} region than for {\tt TestA}.
At the density maximum, the H$_2$ abundance also reaches the maximum value $8.84\E{-5}$
(Fig. \ref{fig:rad_1stP3m_0031}d).
Then, $\abN{H_2}$ is overestimated (Fig. \ref{fig:rad_1stP3m_0031}e), resulting in smaller $\kdiss$ (Fig. \ref{fig:rad_1stP3m_0031}f).
The H$_2$ fraction is still smaller than the threshold for star formation,
and thus secondary star formation does not occur for {\tt TestB}.

%%%%%%%%%%%%%%%%%%%%%%%%%%%%%%%%%%%%%%%%%%%%%
% Discussion %%%%%%%%%%%%%%%%%%%%%%%%%%%%%%%%%%%%%
%%%%%%%%%%%%%%%%%%%%%%%%%%%%%%%%%%%%%%%%%%%%%
\section{Discussion}
\label{sec:discussion}

\subsection{Computational time}
\label{sec:computational_time}

We carry out the simulations with 448 cores on the Frontera supercomputer system
at Texas Advanced Computing Center.
Table \ref{tab:tcomp} shows the computational time $\tcomp$ for the different
approximate methods and threshold densities $\nHth$ for {\tt M10}.
The computational time is generally longer for {\tt n8} than for {\tt n6} by a factor of
$\sim 10$.
Because the gas density increases up to $10^8~\percc$ in {\tt n8}, it takes additional
computation to solve hydrodynamics, chemistry and radiative transfer
in the region with densities $\sim 10^6 < \nH / \percc < 10^8$.

For {\tt n6}, $\tcomp$ is shorter at 6.35 hours for {\tt TestB} than the 15.5 hours taken
for {\tt TestA} by a factor of 2.5.
The local approximation can reduce the numerical cost, compared to solving the
radiative transfer equation of LW photons.
We estimate the fraction $\fRT$ of computational time for radiative transfer calculation
that includes the ionizing photons and, for {\tt TestA}, the LW photons
(fourth column of Table \ref{tab:tcomp}).
For {\tt TestB}, $\fRT = 46.1$\%, smaller than 86.9\% for {\tt TestA}.
For {\tt TestC}, $\tcomp$ is similar to {\tt TestA}.

In Mpc-scale cosmological simulations, hundreds of Pop III stars form
by a redshift $\sim 10$ \citep{Skinner20, Jeon21, Schauer21}.
It is costly to solve the radiative transfer equation of LW photons for all the stars,
because LW photons reach longer distance than ionizing photons
by two orders of magnitude (see Fig. \ref{fig:rad_1stP3_0031}).
The density gradient approach can reduce the computational time by a factor of $2.5$,
reproducing the star formation history although it fails to include the
contribution of the H$_2$-ring to the column density and thus
underestimates the column density.

\subsection{Feedback effects}

We have studied that the approximation methods
of LW radiative transfer and their effects onthe multiplicity of Pop III stars 
in a MH.
This may significantly affect the efficiency of radiative and SN feedback.
In this section, we discuss the impact from the different numerical
setups in a quantitative manner.

\subsubsection{Ionization feedback}

The emission rates of UV photons is roughly proportional to the number of massive Pop III stars. 
In Fig. \ref{fig:prof_0048}a, we compare the radial profiles of 
the H$^+$ and He$^+$ abundances at the time $\tSF = 0.1$ Myr 
for the three approximation methods for {\tt n6}.
We define the radius of the I-front as the distance where
$\aby{H^+} = 0.01$, which is comparable to
the radius of the I-front for He.
The I-front reaches $31.3$ comoving pc for {\tt TestC},
which is larger than $\sim 5$ pc for the other runs,
because multiple radiation sources form.

In this work, we terminate the simulations at $\tSF = 0.1$ Myr,
but we can predict whether the {\HII} region will eventually expand
beyond the virial radius.
%and reach the intergalactic medium (IGM) for {\tt TestC}.
\citet{Chiaki18} estimated the critical halo mass, below which
the radiation energy exceeds the binding energy of a MH, to be
\begin{eqnarray}
M_{\rm halo, cr} &=&
5.64\E{6} \ \Msun
\left( \frac{v_{\rm D}}{10 \ {\rm km \ s^{-1}}} \right) ^{3/4} 
\left( \frac{\tlife}{10 \ {\rm Myr}} \right) ^{3/4} \nonumber \\ && \times
\left( \frac{\QHI}{5\E{49} \ {\rm s^{-1}}} \right) ^{3/4}
\left( \frac{1+z}{26} \right) ^{-3/2},
\label{eq:MHion}
\end{eqnarray}
where $v_{\rm D}$ is the expansion velocity of a D-type front, and
$\tlife$ is the stellar lifetime.
A Pop III star with a mass $10.4~\Msun$ emits ionizing photons
at a rate $5.42\E{47}~{\rm s}^{-1}$ (Table \ref{tab:stars}).
In models where a single star forms ({\tt TestA} and {\tt B}),
the critical halo mass is $1.89\E{5}~\Msun$.
Since the mass of the host halo $3.39\E{5}~\Msun$ exceeds the critical mass,
we can predict that the {\HII} region does not expand beyond the virial radius.
For {\tt TestC}, the total emission rate is $3.88\E{49}$ and $6.53\E{49}~{\rm s}^{-1}$
for {\tt n6} and {\tt n8}, respectively.
Since the halo mass is below the critical mass ($4.66\E{6}$ and $6.89\E{6}~\Msun$,
respectively), we can predict that
ionization photons can reach IGM.
This indicates that the different models can affect the initial stage of cosmic reionization.

\begin{table}
\begin{minipage}{\columnwidth}
\caption{Computational time for each run}
\label{tab:tcomp}
\begin{tabular}{ccccc}
\hline 
$\MPopIIIp$ &  $\nHth$ & Test & $\tcomp$ & $\fRT$ \\
 $[\Msun ]$ &  $[\percc]$ &      &  [hours] & [\%] \\
\hline
$10.4$      & $10^6$     &  A   & 15.5   & 86.7  \\
            &            &  B   & 6.35   & 46.1  \\
            &            &  C   & 16.4   & 36.9  \\
\hline
$10.4$      & $10^8$     &  A   & 143    & 86.9  \\
            &            &  B   & 65.1    & 24.8 \\
            &            &  C   & 59.9    & 48.0 \\
\hline
\end{tabular}
\medskip \\
Note --- 
$\tcomp$: Computational time in units of hour.
$\fRT$: Fraction of computational time for radiative transfer.
We carry out these simulations with 448 cores on the supercomputer
system TACC/Frontera.
\end{minipage}
\end{table}

\subsubsection{LW feedback}

LW radiation can suppress star formation in neighboring clouds
by dissociating H$_2$ \citep{OShea08, Hirano15} or
sometimes trigger the formation of supermassive stars and
black holes \citep{Omukai01, Wise19, Regan20}.
We compare the LW intensity
\begin{equation}
J_{\rm LW} = \frac{\fsh }{4\pi} \frac{E_{\rm LW}}{\Delta \nu _{\rm LW}} \frac{\QLW}{4 \pi D^2 } ,
\end{equation}
where $D$ is the distance from a star (cluster),
and $\Delta \nu _{\rm LW} = 5.80\E{14}$ Hz is the width of the LW band.
Hereafter we use the LW intensity $J_{21}$ in units of 
$10^{-21}$ erg s$^{-1}$ cm$^{-2}$ Hz$^{-1}$ sr$^{-1}$.

\begin{figure}
\includegraphics[width=\columnwidth]{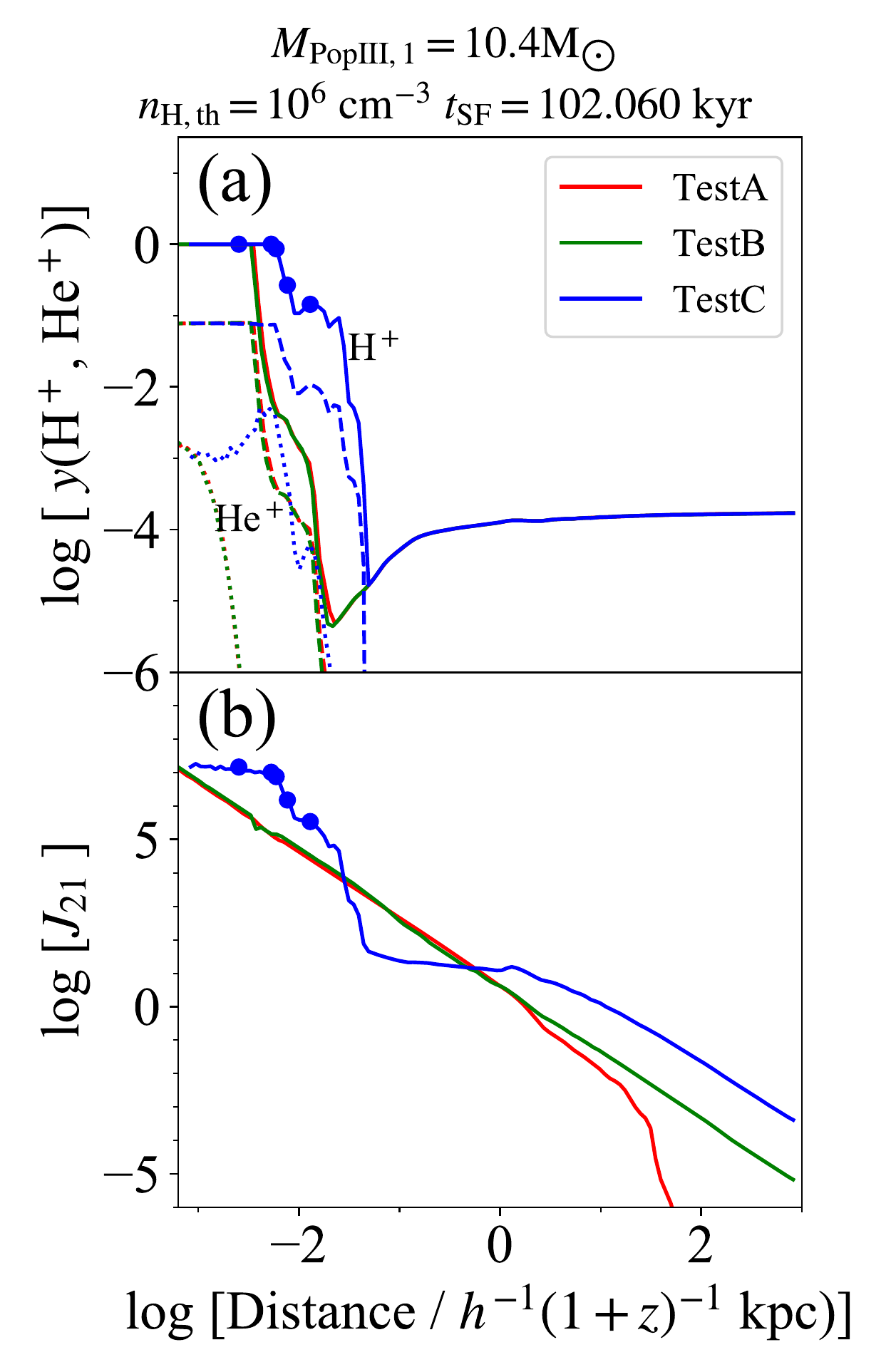}
\caption{
(a) Abundances of H$^+$ (solid curves) and He$^+$ (dashed curves)
relative to hydrogen nuclei and
(b) LW intensity $J_{21}$
in units of $10^{-21}$ erg s$^{-1}$ cm$^{-2}$ Hz$^{-1}$ sr$^{-1}$
as a function of distance from the primary
Pop III star at 100 kyr after the primary star formation.
We take the average in each radial bin.
The red, green and blue curves indicate the results for {\tt TestA}, {\tt TestB}
and {\tt TestC}, respectively.
The blue circles denote the positions of the secondary stars for {\tt TestC}.
}
\label{fig:prof_0048}
\end{figure}

Fig. \ref{fig:prof_0048}b shows the LW intensity as a function of
the distance from the primary star at the time $\tSF = 0.1$ Myr.
At distances $D \lesssim 0.01$ kpc (comoving), the intensity is largest for {\tt TestC}
because of multiple Pop III star formation.
At $0.01 \lesssim D / {\rm kpc} \lesssim 1$, the intensity for {\tt TestC} is smaller
than in the optically thin case, where the intensity declines as $\propto D^{-2}$
due to geometrical dilution.
This indicates that dense clumps absorb LW photons, and additional star formation
may occur in the self-shielded regions for {\tt TestC}.
At $D \gtrsim 1$ kpc, the intensity roughly follows a profile $\propto D^{-2}$
for all the tests, but the intensity is the largest for {\tt TestC}.
LW radiation can quench star formation in low-mass MHs with intensities 
$J_{21} \gtrsim 0.1$ \citep{OShea08}.
$J_{21}$ exceeds this value in larger region of $3.84$ kpc in {\tt TestC}.
Star formation may be delayed for $\sim$ Myr in neighboring MHs
due to strong LW emission from the star cluster.

At $D\sim 10$ kpc, the intensity rapidly declines for {\tt TestA}, 
because the gas is optically thick in the LW band.
For {\tt TestB}, the gas remains optically thin, and the profile follows $\propto D^{-2}$.
The local approximation with density gradient can safely estimate the
column density in the D-front, but it fails to estimate the column density
at larger distances $D\gtrsim 30$ kpc.
Fortunately, at these distances, $J_{21}$ is not so strong ($\sim 10^{-2}$),
and thus star formation may not be significantly affected
by this overestimate of $J_{21}$.

\subsubsection{SN feedback}

Massive Pop III stars will undergo SN explosions at the end of their lives
and release the first metals into ISM and IGM.
\citet{Chiaki18} found that there are two modes of metal enrichment:
internal enrichment (IE) and external enrichment (EE).
The latter occurs if the emission rate of ionizing photons
is sufficiently large before SN explosions occur.
If an {\HII} region expands beyond the virial radius of a host halo, SN shocks
can propagate through the rarefied region 
without energy loss by radiative cooling.
Therefore, EE occurs for a halo with masses below the critical value of
Eq. (\ref{eq:MHion}).

For {\tt TestA} and {\tt B}, we can predict that, since the {\HII} region will be bound
in the MH, IE will occur.
The ejected metal mass is $\Mmet \sim 1~\Msun$ for a progenitor
mass $\MPopIII \sim 10~\Msun$ \citep{Nomoto13}.
If the metals are uniformly mixed with the pristine gas with a
mass $\Mgas = 5\E{4}~\Msun$ in the MH, the
metallicity is estimated to be
\begin{equation}
Z_{\rm IE} = \frac{\Mmet}{\Mgas} = 10^{-3}~\Zsun 
\left( \frac{\Mmet }{1~\Msun} \right) 
\left( \frac{\Mgas }{5\E{4}~\Msun} \right) ^{-1} 
\end{equation}
as confirmed by earlier numerical studies \citep{Ritter12, Ritter15, Sluder16, Chiaki19}.
For {\tt TestC}, since the halo mass is below the critical mass, 
the cloud will be disrupted by multiple SN explosions.
SN ejecta will reach neighboring halos.
However, only a small fraction of ejecta can reach the center of the halos
because of the pressure gradient force of gas clumps.
Therefore, EE is expected to be less effective than IE.
The resulting metallicity will be $Z_{\rm EE} \lesssim 10^{-5}~\Zsun$
in the neighboring halos \citep{Smith15, Chen17, Chiaki18}.

\subsection{Applicability to other astrophysical problems}

We have found that the Sobolev approach is the most
accurate approximation in the triggered star formation scenario.
%In general, the Sobolev approach is valid when the velocity gradient is monotonic over
%the Sobolev length.
%Although the velocity gradient fluctuates in our cases (Figs. \ref{fig:rad_1stP3_0031}e, 
%\ref{fig:rad_1stP3h_0035}e and \ref{fig:rad_1stP3m_0031}e), the fluctuations are not so significant 
%to alter the star formation history in the shell.
There are other astrophysical problems which can be affected by the
different approximation methods of LW transfer.
In this section, we discuss that our finding can be applied to other problems.

When the first Pop III stars form, H$_2$ 
line cooling plays a crucial role in the fragmentation 
of primordial collapsing clouds \citep{Bromm99, Abel02, Yoshida03}.
\citet{Greif14} studied the escape fraction of photons in the transition lines, 
comparing a full ray tracing model and the Sobolev approximation.
They ran high-resolution simulations which can resolve small-scale turbulence.
They found that the escape fraction is overestimated with the Sobolev approximation, 
because the length scale $\lshV$ of velocity fluctuations is too small
compared to the bulk inflow velocity.

Direct collapse is one of the possible pathways to SMBH formation 
\citep[][for a review]{Inayoshi20}. 
If a primordial halo is exposed by sufficiently strong LW background, H$_2$ molecules
are dissociated.
The cloud can collapse directly to a supermassive star that forms a massive BH seed, avoiding fragmentation.
LW transfer is crucial to determine the critical LW intensity above which direct collapse can occur.
\citet{WolcottGreen11} found that the Sobolev approach can well reproduce the result of
a full ray-tracing calculation.
This indicates that the Sobolev approach is applicable to the direct collapse scenario.

Terrestrial exoplanets can lose their atmospheres when they are exposed 
strong X-ray and extreme ultraviolet (EUV) irradiation from M dwarf host stars \citep{Luger15}.
The mass loss rate can be suppressed by molecular cooling (e.g., H$_2$ and H$_2$O ro-vibrational
transition line cooling),
and the cooling efficiency depends on the treatment of radiation transfer.
In their atmospheres, the velocity gradient changes at length scales comparable to or less than
the Sobolev length of $\sim 1$--$10$ times planet radius \citep{Yoshida22}.
Therefore, the Sobolev approximation cannot be applied to this problem.

\subsection{Caveats}

\subsubsection{Multifrequency effects}
\label{sec:multifrequency}

In our direct integration model,
we have not considered the frequency dependence of the photodissociation cross-section.
The cross-section has spikes at frequencies corresponding to 
resonance lines \citep[e.g.,][]{Heays17}, and Doppler shifts of the lines can cause
photon escape in a fluid moving relative to a source \citep{WolcottGreen11}.
In the context of Pop III star formation, \citet{Greif14} evaluated this effect 
with their multifrequency radiation transport model.
They found that the escape probability of H$_2$ lines are underestimated by a factor of two 
if the Doppler shift is not considered.
We can expect that the effect can be important also in the case of triggered star formation.
In our simulations, the radial velocity of the {\HII} shell ($\sim 20~\kmpers$) is larger than 
the thermal velocity 
\begin{equation}
v_{\rm th} = \left( \frac{kT}{2\mH} \right) ^{1/2} = 5~\kmpers \left( \frac{T}{5000~{\rm K}} \right)^{1/2}.
\end{equation}
Thus, a larger fraction of dissociation photons can escape if we had considered the multifrequency effect,
possibly leading to secondary star formation being suppressed further.

\subsubsection{Updated shielding function}

We have used the shielding function $\fsh (\abN{H_2})$ presented
by \citet{WolcottGreen11}.
\citet{WolcottGreen19} lately updated the shielding function, including the effect
of non-local thermal equilibrium (LTE) populations of H$_2$ molecules.
In the D-type front, the density, temperature and LW intensity are typically
$\nH \sim 10^5~\percc$, $T\sim 5000$ K and $J_{21} \sim 10^3$, respectively, where
the non-LTE effect is not negligible.
In this regime, we have underestimated the shielding factor.
If we use the updated shielding function, $\fsh (\abN{H_2})$ will be larger, and 
there will be more chance to suppress secondary star formation.
This strengthens our conclusion in the more accurate methods: direct integration
and the Sobolev approximation.
In any case, we will include the non-LTE effect for more physically-motivated formulation
of the shielding function.

\subsubsection{HD photodissociation}

In this work, we do not consider photodissociation of
HD molecules, another important coolant in 
collapsing gas clouds \citep{Johnson06, Hirano14}.
HD molecules absorb photons in the LW band but in lines 
at different frequencies from H$_2$.
The gas is generally optically thin in the absorption lines, because
the HD abundance is smaller than H$_2$ by 
five orders of magnitude \citep{Omukai12}.
In our implementation, we assume that each photon package is 
monochromatic, and we use the shielding function (Eq. \ref{eq:fsh}) 
averaged over all frequencies in the LW band \citep{WolcottGreen11}.
HD molecules % share the same photon type with H$_2$ and
could receive only a fraction of photons that were not 
absorbed by H$_2$.
To overcome this problem, it is ideal to separate the photon package 
into three energy bins that interact with H$_2$, HD and both.
We will improve our model in forthcoming papers.

%\subsubsection{Stellar mass}
%
%We have fixed the primary star mass to $10.4~\Msun$.
%The density and temperature structure of an {\HII} region should change
%for different stellar masses
%\citep{Kitayama04, Whalen08}.
%For a more massive primary star, the density increases in a D-type front
%more rapidly because of the more intense ionizing photon emission.
%At the same time, the emission rate of LW photons is also larger,
%which may suppress the formation of the secondary stars. 
%We will study star formation for a wider range of stellar masses in future work.

%%%%%%%%%%%%%%%%%%%%%%%%%%%%%%%%%%%%%%%%%%%%%
% CONCLUSION %%%%%%%%%%%%%%%%%%%%%%%%%%%%%%%%%%%%%
%%%%%%%%%%%%%%%%%%%%%%%%%%%%%%%%%%%%%%%%%%%%%
\section{Conclusion}
\label{sec:conclusion}

Massive stars emit tremendous amounts of ionizing photons,
creating an {\HII} region.
At the limb of the {\HII} region, a dense shell forms
due to internal thermal pressure.
This D-type front is a potential star-forming site 
\citep[triggered star formation;][]{Elmegreen77, Whitworth94, Hosokawa05, Hosokawa06}.
In this work, we find that star formation in the D-type front depends on 
the numerical scheme to solve LW radiation transport.
The LW flux depends on the estimate of the H$_2$ column density $\abN{H_2}$.
We test three cases: the direct integration of H$_2$ density ({\tt TestA}),
local approximation based on the density gradient ({\tt TestB}) and the Jeans length ({\tt TestC}).

We compare the number of secondary stars forming in the D-type front.
No secondary stars form in {\tt TestA} and {\tt B} while
five stars form in {\tt TestC}.
In {\tt TestA}, dissociating photons are only partially ($\sim 0.3$) absorbed 
in a thin H$_2$-ring, 
and the secondary star formation is suppressed.
In {\tt Test B}, the result is consistent with {\tt TestA}, but we caution
that the local approximation underestimates or overestimates the column density
when the primary Pop III stellar mass is $10$ and $40~\Msun$, respectively.
In {\tt TestC}, the number of forming stars is overestimated
because the Jeans length is generally larger than the thickness of the H$_2$-ring.

It is numerically expensive to solve radiation transport of LW photons in numerical simulations
because the gas is typically optically thin in the LW band, and photons
can reach a large distance ($\sim 10$ comoving kpc).
In large-volume cosmological simulations with a side of $\sim$ comoving Mpc, 
the local approximation is useful to reduce computational costs.
We find that the computational time is reduced for the local approximation
with the density gradient ({\tt TestB}) by a factor of 2--3, compared to 
direct integration of $\abN{H_2}$ ({\tt TestA}).
Although the local approximation has limitations, 
the density gradient approach is balanced
strategy to reproduce the star formation history in the early stage of
structure formation
while keeping computational costs low.

%%%%%%%%%%%%%%%%%%%%%%%%%%%%%%%%%%%%%%%%%%%%%%
% ACKNOULEDGEMENTS %%%%%%%%%%%%%%%%%%%%%%%%%%%%%%%%
%%%%%%%%%%%%%%%%%%%%%%%%%%%%%%%%%%%%%%%%%%%%%%
\section*{ACKNOWLEDGMENTS}

We thank Greg Bryan, Zoltan Haiman, Tatsuya Yoshida and Yuichi Ito for helpful discussion.
GC is supported by Research
Fellowships of the Japan Society for the Promotion of Science (JSPS). 
JHW is supported by National Science Foundation grants OAC-1835213 and 
AST-2108020 and NASA grants NNX17AG23G, 80NSSC20K0520, and 80NSSC21K1053.
The simulation was performed
with NSF's LRAC allocation AST-20007 on the Frontera
resources in TACC. 
The figures in this paper are constructed with the
plotting library {\sc matplotlib} \citep{matplotlib}.

\section*{Data availability}

The versions of {\sc enzo}, {\sc grackle}, and {\sc yt}
used in this work are available at 
\url{https://github.com/genchiaki/enzo-dev/tree/metal-dust},
\url{https://github.com/genchiaki/grackle/tree/metal-dust},
\url{https://github.com/genchiaki/yt/tree/metal-dust},
respectively.
The script for {\sc yt} used in this work is available at
\url{https://github.com/genchiaki/Analysis_cosmo}.
The simulation data will be shared on reasonable request to the authors.

%%%%%%%%%%%%%%%%%%%%%%%%%%%%%%%%%%%%%%%%%%%%%%
% APPENDIX  %%%%%%%%%%%%%%%%%%%%%%%%%%%%%%%%%%%%%%
%%%%%%%%%%%%%%%%%%%%%%%%%%%%%%%%%%%%%%%%%%%%%%

%\appendix
%\section{Dust cooling rates with grain growth}

%%%%%%%%%%%%%%%%%%%%%%%%%%%%%%%%%%%%%%%%%%%%%
% REFERENCES %%%%%%%%%%%%%%%%%%%%%%%%%%%%%%%%%%%%
%%%%%%%%%%%%%%%%%%%%%%%%%%%%%%%%%%%%%%%%%%%%%

\label{lastpage}

\end{document}